\documentstyle[twoside,fleqn]{article}

\newcommand{\bls}[1]{\renewcommand{\baselinestretch}{#1}}
\def\noi{\noindent}

\makeatletter
\renewcommand{\section}{\@startsection{section}{1}{0pt}%
        {-3.5ex plus -1ex minus -.2ex}{2.3ex plus .2ex}%
        {\large\bf\protect\raggedright}}

\renewcommand{\subsection}{\@startsection{subsection}{2}{0pt}%
        {-3ex plus -1ex minus -.2ex}{1.4ex plus .2ex}%
        {\normalsize\bf\protect\raggedright}}

\renewcommand{\thesubsubsection}%
        {\arabic{section}.\arabic{subsection}.\arabic{subsubsection}.}
\renewcommand{\@oddhead}{\raisebox{0pt}[\headheight][0pt]{%
   \vbox{\hbox to\textwidth{\rightmark \hfil \rm \thepage \strut}\hrule}}}
\renewcommand{\@evenhead}{\raisebox{0pt}[\headheight][0pt]{%
   \vbox{\hbox to\textwidth{\thepage \hfil \leftmark \strut}\hrule}}}
\newcommand{\heads}[2]{\markboth{\protect\small\it #1}{\protect\small\it #2}}
\newcommand{\Acknow}[1]{\subsection*{Acknowledgement} #1}
\makeatother

\newcommand{\Title}[1]{\noi {\Large #1} \\}
\newcommand{\Author}[2]{\noi{\large\bf #1}\\[2ex]\noi{\it #2}\\}
\newcommand{\Abstract}[1]{\vskip 2mm \begin{center}
     \parbox{16.4cm}{\small\noi #1} \end{center}\bigskip}
\newcommand{\foom}[1]{\protect\footnotemark[#1]}

\newcommand{\email}[2]{\footnotetext[#1]{e-mail: #2}}

\def\nq{\hspace{-1em}}
\def\nqq{\hspace{-2em}}
\def\nhq{\hspace{-0.5em}}

\def\cm{\hspace{1cm}}

\def\Ref#1{Ref.\,\cite{#1}}

\newcommand{\Eq}[1]{Eq.\,(\ref{#1})}

\def\eqs{Eqs.\,}
\def\beq{\begin{equation}}
\def\eeq{\end{equation}}
\def\bear{\begin{eqnarray}}
\def\al{&\nhq}
\def\lal{&&\nqq {}}               
\def\bearr{\begin{eqnarray} \lal}
\def\ear{\end{eqnarray}}
\def\earn{\nonumber \end{eqnarray}}
\def\dst{\displaystyle}
\def\tst{\textstyle}

\newcommand{\fract}[2]{{\tst\frac{#1}{#2}}}
\def\nn{\nonumber\\ {}}

\def\nnn{\nonumber\\ \lal }

\def\yy{\\[5pt]}

\def\eql{\al =\al}

\def\eqdef{\stackrel{\rm def}{=}}
\def\e{{\,\rm e}}
\def\d{\partial}

\def\sign{\mathop{\rm sign}\nolimits}
\def\diag{\mathop{\rm diag}\nolimits}
\def\dim{\mathop{\rm dim}\nolimits}
\def\const{{\rm const}}
\def\Half{{\dst\frac{1}{2}}}
\def\half{{\tst\frac{1}{2}}}
\def\then{\ \Rightarrow\ }

\newcommand{\vars}[1]{\left\{\begin{array}{ll}#1\end{array}\right.}
\newcommand{\lims}[1]{\mathop{#1}\limits}

\def\wider{\vphantom{\int}}

\def\m{{\rm m}}
\def\sumt{\sum_{i=1}^{3}}

\newcommand{\R}{{\sf R\hspace*{-0.9ex}\rule{0.1ex}{1.5ex}\hspace*{0.9ex}}}

\def\phim{\mbox{$\phi^{\min}$}}
\newcommand{\T}[1]{\raisebox{0.2ex}{$\lims{T}_{#1}$}{}}
\newcommand{\TAB}[1]{\raisebox{0.2ex}{$\lims{T}_{#1}$}{}_A^B}
\def\mO{\mbox{[-1]}}
\def\pO{\mbox{[1]}}

\def\mn{_{\mu\nu}}

\def\sph{spherically symmetric\ }
\def\bh{black hole}

\heads{Kirill A. Bronnikov}
{Multidimensional Cosmology with a Generalized Maxwell Field: Integrable
Cases}

\bls{1.02}

\begin{document}
\thispagestyle{empty}
\twocolumn[
\noi \unitlength=1mm
\begin{picture}(174,8)
       \put(31,8){\shortstack[c]
       {RUSSIAN GRAVITATIONAL SOCIETY\\
       INSTITUTE OF METROLOGICAL SERVICE \\
       CENTER FOR GRAVITATION AND FUNDAMENTAL METROLOGY }        }
\end{picture}
\begin{flushright}
                                         RGS-VNIIMS-97/03\\
                                         gr-qc/9704012 \\
     	          {\it Grav. and Cosmol.} {\bf 3}, 1(9), 65--70 (1997)
\end{flushright}
\medskip

\Title{MULTIDIMENSIONAL COSMOLOGY\yy
	  WITH A GENERALIZED MAXWELL FIELD: \yy
	  INTEGRABLE CASES}

\Author{Kirill A. Bronnikov\foom 1}
{Centre for Gravitation and Fundamental Metrology, VNIIMS,
     3--1 M. Ulyanovoy St., Moscow 117313, Russia}

\Abstract
{We consider multidimensional cosmologies in even-dimensional space-times
($D=2n$) containg perfect fluid and a multidimensional generalization of the
Maxwell field $F_{A_1\cdots A_n}$ preserving its conformal invariance.
Among models with an isotropic physical 3-space some integrable cases
are found: vacuum models (which are integrable in the general case) and
some perfect fluid models with barotropic equations of state. All of them
contain a component of the $F$ field appearing as an additional scalar in 4
dimensions. A two-parameter family of spatially flat models and four
one-parameter families, including non-spatially flat models, have been
obtained (where the parameters are constants from the fluid equation of
state). All these integrable models admit the inclusion of a massless scalar
field or an additional fluid with the maximally stiff equation of state.
Basic properties of vacuum models in the physical conformal frame are
outlined.
}

] 
\email 1 {kb@goga.mainet.msk.su}

\section{Introduction}

	This paper continues a study of exact solutions to the
	gravitational field equations including a generalization of the Maxwell
	field to $2n$-dimensional space-times ($n=2,3,...$), began in
	\Ref{122}.  Unlike the conventional straightforward use of the Maxwell
	field in $D$ dimensions, this generalization preserves the invariance
	of the field under multidimensional conformal transformations.

     This generalization (Generalized Maxwell Field, GMF) is represented by
     a $(D/2)$-form, where $D=2n$ is the space-time dimension:
\beq                                                        \label{dU} 
	F = dU,\qquad {\rm or} \qquad
	F_{A_1\ldots A_n} = n! \d_{\,[A_1} U_{\,A_2\ldots A_n]}
\eeq
     where $U$ is a potential $(n-1)$-form and square brackets denote
     alternation.  The field $F$ is invariant with respect to the gauge
     transformation
\beq
	U \ \mapsto \ U + dZ                                   \label{dZ} 
\eeq
     where $Z$ is an arbitrary $(n-2)$-form; in 4 dimensions this is the
     conventional gradient transformation of the Maxwell field.

	The GMF was introduced in Refs.\,\cite{Fa1,Fa2}, where
	some particular cosmological models with the GMF were studied; in
	\cite{122} we discussed in some details several families of \sph\
	solutions, with combinations of electric, magnetic and quasiscalar type
	components of the GMF, with possible inclusion of a massless, minimally
	coupled scalar field $\varphi = \phim$. Some cases of interest were
	found, namely, \bh s with a somewhat unusual mass and charge dependence
	of the Hawking temperature.

	This paper presents some classes of exact cosmological solutions,
	including those of \cite{Fa1,Fa2,Fa3} as special cases. Just as in
	those papers, only Friedmann-like models, isotropic in the three
     physical dimensions, are considered, therefore the GMF can have only
	``quasiscalar" components, behaving as scalars in the physical
	space-time. The solutions found here contain only one such component.
	Further generalizations with a number of such components and even a
	number of $p$-forms, as discussed in field models like M-theory
     (see e.g. \cite{M} and references therein), are possible.

	The aim of this paper is basically to indicate the integrable cases. A
	more detailed description of the solutions will be given elsewhere.

	Only the behaviour of vacuum solutions is briefly outlined, some of
	them exhibiting attractive features. In particular, solutions with
	a hyperbolic 3-space, have a linearly expanding physical space at their
	late-time asymptotic, with internal-space scale factors tending to
     finite constant values which (bearing in mind the usual assumption
     that the internal spaces are compact) may be arbitrarily small.

	Throughout the paper capital Latin indices range over all $D$
	space-time dimensions, Greek ones take the values 0,1,2,3, indices like
	$a_1,\ldots,a_n$ refer to internal spaces and $i,j$ enumerate the
	factor spaces.

\section
	{Generalized Maxwell field and 3-isotropic cosmology.\protect\newline
	Field equations}                                            

     Consider general relativity in a Riemannian space-time $V^D$ ($D=2n$,
     $n\geq 2$), in the presence of the GMF and perfect fluid (matter),
	with the total action
\bearr
 S = \int  d^D x\sqrt{g}\bigr[R/(2\kappa) +(-1)^{n-1}F^2 + L_\m]     
                                    \nnn                       \label{a}
     \cm     F^2 \equiv F^{A_1...A_n}F_{A_1...A_n},
\ear
     where $R={}^DR$ is the scalar curvature corresponding to the $D$-metric
     $g_{AB}$, $g=|\det g_{AB}|$, $\kappa$ is the $D$-dimensional
	gravitational constant, $F$ is the GMF and $L_\m$ is the matter
	Lagrangian.

     The $F$-sector of the action is invariant under conformal
     mappings of $V^D$ ($g_{AB} \mapsto f(x^A)g_{AB}$) provided the
	components $F_{A_1\ldots A_n}$ are unchanged in such transformations
     (the potential $U$ from (\ref{dU}) may
	experience a gauge transformation (\ref{dZ}) where $Z$ is
	some $(n-2)$-form depending on the conformal factor $f$).

     The field equations are                        
\bear
	G^B_A \equiv R^B_A - \half \delta^B_A R \eql -\kappa T^B_A,
                                                           \label{eqE}\yy
                 \nabla_A F^{AA_1\ldots A_n} \eql 0,       \label{eqF}
\ear
    where the energy-momentum tensor (EMT) $T^B_A$ is a sum of contributions
	from $F$ and matter. The EMT of the $F$ field is
\beq                                                     \wider    
	\TAB{F} = (-1)^{n-1}\bigl( nF_{AC_2...C_n}F^{BC_2\ldots C_n} -
       \half \delta^B_A F^2\bigr).                             \label{EMT-F}
\eeq
	An alternative form of (\ref{eqE}) is
\beq                                                               
	R^B_A = -\kappa \TAB{F}
	        -\kappa \biggl(\TAB{m}-\frac{1}{D-2}\delta_A^B\T{m}_C^C\biggr)
												   \label{c10}
\eeq
     where $\TAB{m}$ is the EMT of all matter except the $F$ field.

	In what follows we will assume that the physical 3-space is isotropic
    and the GMF has only one (up to index permutations) nontrivial component
	compatible with such isotropy, namely,
\beq
	F_{0a_2\ldots a_n},                                     \label{F1}
\eeq
	where the index 0 corresponds to time and all the others to extra
	dimensions; moreover, we restrict ourselves to the simplest
	space-time structure compatible with such an $F$ field. Evidently,
	among the $2n-4$ internal dimensions \Eq{F1} distinguishes two
	subsets, with and without the $F$ field, or, in other words, $n-1$
	coordinates from the remaining $n-3$.  Thus for $n>3$ there must be at
	least two internal factor spaces with different dynamics.

	Accordingly, we consider $V^{2n}$ with the structure
\beq                                                    	\label{Stru}
     V^D = \R \times M_1 \times M_2 \times M_3,
\eeq
     where $\R$ corresponds to time, $M_1$ is the external isotropic 3-space
     (of curvature $K = 0, \pm 1$), $M_2$ and $M_3$
	are Ricci-flat internal spaces, $\dim M_2=n-1$, $\dim M_3 = n-3$, so
	that $M_2$ is parametrized by the coordinates whose numbers belong to
	$a_k$ in (\ref{F1}) and $M_3$ to the remaining ones.

	The metric is assumed in the form
\beq     \wider
	ds^2 = \e^{2\gamma}d\tau^2 - \e^{2\beta_1}ds_1^2
	     - \e^{2\beta_2}ds_2^2 - \e^{2\beta_3}ds_3^2          \label{c6}
\eeq
	where $\gamma,\ \beta_i$ are functions of  the time variable $\tau$
	and $ds_i$ are the $\tau$-independent metrics in $M_i$.

	For the perfect fluid we assume a pressure isotropic in each
     factor space, so that its EMT is
\beq           \wider                                           \label{c7}
	\TAB{\m} = \diag (\rho,\ [-p_1]_3,\ [-p_2]_{n-1}\ [-p_3]_{n-3})
\eeq
	where $\rho$ is the density, $p_i$ are the pressures in the
	corresponding directions and the symbol $[x]_k$ means that a quantity
	$x$ occupies $k$ positions along the diagonal. We admit that $\rho$ and
	$p_i$ are connected by the equations of state
\beq
	p_i = (a_i - 1)\rho, \cm  a_i = \const,                  \label{c8}
\eeq
	so that a physical range of $a_i$ is between $0$ (vacuum-like state)
	and 2 (stiff matter).

	It should be noted that the 4-dimensional part of the $D$-metric
	(\ref{c6}) does not coincide with the metric to be used to
	interpret actual cosmological observations. Indeed,
     real space-time measurements (such as Solar-system experiments or
	cosmological redshift measurements) rest on the constancy of atomic
     quantities (the {\sl atomic system of measurements}). Thus, the modern
	definition of reference length is connected with a certain spectral
	line, determined essentially by the Rydberg constant and ultimately by
	the electron and nucleon masses. Therefore it is reasonable to
	apply as a {\bf physical} one such a metric (to be denoted 	$g^*\mn$)
	conformally related to $g\mn$ with which masses of bodies of
	nongravitational matter, like atomic particles, do not change from
	point to point. This happens if the matter Lagrangian appears in the
	total Lagrangian in 4 dimensions without any $x^\mu$-dependent factor.

	The choice of $g^*\mn$ depends on how $L_m$ appears in
	the original action. In our case (\ref{a}), just as in \Ref{017},
	it is easy to check that the metric in the ``atomic gauge''
	should have the form
\beq
     g^*\mn = \e^{\sigma/2}g\mn                              \label{MapA}
\eeq
	where $\e^\sigma$ is the volume factor of extra dimensions, in our case
	$\sigma \equiv (n-1)\beta_2 + (n-3)\beta_3$. Thus, in particular, the
	physical 3-dimensional scale factor $a(\tau)$ is
\beq
	a(\tau) \equiv \exp\,[\beta_1 + \fract 14 (n-1)\beta_2
                                  + \fract 14 (n-3)\beta_3]. \label{atau}
\eeq

	We are now ready to write down the field equations explicitly. Let us
	use the harmonic time coordinate $\tau$, so that
\beq                                                          \label{c11}
	\gamma(\tau) =  \equiv \sumt N_i \beta_i(\tau),
	                                             \cm N_i = \dim M_i.
\eeq

	The Maxwell-like equation (\ref{eqF}) for the only nonzero GMF
	component $F_{04\ldots n+2}$ is easily integrated giving
\beq
	F^{045\ldots} = q/\sqrt{g}, \cm  q=\const                \label{c12}
\eeq
	where due to (\ref{c11}) $\sqrt{g} = \e^{2\gamma}$. So the EMT of the
	$F$ field may be written as
\bearr                                                         
	\TAB{F} = \half n!\, q^2 \e^{-6\beta_1-2(n-3)\beta_3} \nnn
     \cm \times\diag (1, \mO_3, \pO_{n-1}, \mO_{n-3}).        \label{c13}
\ear

	For perfect fluid the conservation law $\nabla_B \TAB{F}=0$ after
	integration leads to
\beq
	\rho = \rho_0 \exp\biggl(- \sumt N_i a_i \beta_i \biggr),
	\qquad \rho_0 = \const                                   \label{c14}
\eeq
	and its EMT is presented in the form
\bearr                                                        \label{c15}
	\TAB{\m} = \rho_0\e^{-2\gamma+2y}   \nnn
	\cm\times \diag (1,[1-a_1]_3, [1-a_2]_{n-1}, [1-a_3]_{n-3})\nnn
\ear
	where we have introduced the function
\beq
	2y = \sumt (2-a_i)N_i\beta_i.                            \label{c16}
\eeq

	With (\ref{c6}) and (\ref{c11}), the nonzero components of the Ricci
	tensor are
\bear                                                         \label{c17}
	R^\tau_\tau \eql \e^{-2\gamma}
	      \biggl(\ddot\gamma-\dot\gamma^2 + \sumt N_i\beta_i^2\biggr),\\
												  \label{c18}
	R_{m_i}^{n_i} \eql \delta_{m_i}^{n_i}
		        \Bigl[\e^{2\beta_i}K_i(N_i-1)
			                     + \e^{-2\gamma} \ddot \beta_i
\ear
	where $K_i=0, \pm 1$ are the factor space curvatures; we will assume
	$K_2=K_3=0$ and only for the physical space treat all the three
	variants, corresponding to different Friedmann-Robertson-Walker models.

	The spatial components of the Einstein equations now read:
\bear                                                          \label{c19}
	2K_1\e^{2\gamma=2\beta_1}+\ddot \beta_1
	              \eql \quad Q^2 \e^{2x} + A_1 \kappa\rho_0 \e^{2y},\\
												   \label{c20}
	\ddot \beta_2 \eql {} - Q^2 \e^{2x} + A_2 \kappa\rho_0 \e^{2y},\\
				     					     	   \label{c21}
	\ddot \beta_3 \eql \quad Q^2 \e^{2x} + A_3 \kappa\rho_0 \e^{2y},
\ear
	where
\bear                                                          
	x   \al \equiv \al (n-1) \beta_2,                \nn
	A_i \al\eqdef\al a_i - \frac{A-2}{2n-2}, \cm
	                A \eqdef \sumt N_i a_i,\nn  \label{c22}
	Q^2 \eql \half n!\, \kappa q^2.
\ear

	\eqs (\ref{c19})--(\ref{c21}) have a first integral (the energy
	integral), coinciding with the ${\tau \choose \tau}$ component of the
	Einstein equations:
\bearr                                                           
    \nhq 3K_1 \e^{2\gamma-2\beta_1} + \Half \dot\gamma^2
		- \Half \sumt N_i \dot\beta_i^2                  \label{c23}
                 = Q^2 \e^{2x} + \kappa\rho_0 \e^{2y}. \nnn
\ear

	In what follows we try to solve \eqs (\ref{c19})--(\ref{c21});
	(\ref{c23}) is used to obtain an additional relation between the
	integration constants and simultaneously to verify the correctness of
	the solutions.

\section{Vacuum solutions with the $F$ field}

	In the case $\rho_0=0\ \then \rho\equiv p_i \equiv 0$ one easily solves
	\Eq{c20}:
\beq                                                             
	\e^{-(n-1)\beta_2}= \frac{1}{h} \sqrt{(n-1)Q^2}
	                          \cosh [h(\tau-\tau_2)]       \label{c24}
\eeq
	with the integration constants $h > 0$ and $\tau_2$. As now
	$\ddot\beta_2 + \ddot \beta_3 =0$, $\beta_3$ is also easily found:
\beq
	\beta_3 = -\beta_2 + b \tau + b_3, \qquad          \label{c24a}
                                               b,\ b_3 = \const.
\eeq
	Lastly, (\ref{c19}) can be combined with (\ref{c20}) and (\ref{c21})
	to give
\beq                                                        \label{c25}
     (\gamma-\beta_1)\,\ddot{} + 4K_1 \e^{2\gamma-2\beta_1}=0,
\eeq
     whence                                                     
\bearr                                                          \label{c26}
 \nq	\e^{-z} \equiv \e^{\beta_1-\gamma} = \vars{
	                     2s(k,\tau),    & K_1 =-1\  (k\in \R);    \\
                          \e^{k\tau},     & K_1=0, \ \  (k\in \R);  \\
                      2k^{-1}\cosh k\tau, & K_1 =1,\ \  (k > 0),     } \nnn
\ear
	where
\beq                                                            
	s(k,\tau) \eqdef \vars { k^{-1}\sinh k\tau, &  k>0; \\     \label{c27}
					          \tau,         &  k=0; \\
					     k^{-1} \sin k\tau, &  k<0;     }
\eeq
	$k=\const$ and one more constant is eleminated by shifting the
	origin of $\tau$. The integration is completed.

	A substitution to the energy integral (\ref{c23})	gives:
\bear                                                           
	3K \eql \frac{4}{n-1}h^2 + (n-1)(n-3)b^2, \nn            \label{v1}
	 K \al \eqdef \al \vars{k^2\sign k, & K = -1;\\
					    k^2,        & K = 0,+1.   }
\ear
	so that, in particular, for $K_1=-1$ we have $k>0$ and
     $ \e^{-z} = (2/k)\sinh k\tau $.

     The physical components of the metric are (see (\ref{MapA}),
     (\ref{atau})).
\bear                                                            
	g^*_{00}  \eql \e^{2\gamma^*} = \exp\,[3z -\beta_2 -\half (n-3)b\tau];
	                                                            \nn
     a^2(\tau) \eql \exp \,[z-\beta_2 -\half (n-3)b \tau].     \label{v2}
\ear

	The model properties strongly depend on the dimension $2n$ and specific
	values of the integration constants. Thus, the cosmic proper time
	$t = \int \exp(\gamma^*)\, d\tau$ is finite or infinite depending on
     the convergence or divergence of the integral, etc. One can conclude
	the following.

\medskip\noi
	{\bf 1.} The harmonic time $\tau \in\R$ for $K_1=0, +1$ and
	(with no generality loss) $ \tau\in (0,\infty)$ for $K_1=-1$.

\medskip\noi
	{\bf 2.} The internal-space scale factor $\e^{\beta_2}$ tends to zero
	as $\tau\to \pm \infty$, i.e. at both ends of the evolution process for
	$K_1=0, +1$ and at one end for $K_1=-1$. In the latter case, $\beta_2$
	tends to a finite limit as $\tau\to 0$.

\medskip\noi
	{\bf 3.} Another scale factor $\e\beta_3$ (existing only for $n>3$)
	tends to infinity either as $\tau\to +\infty$, or as $\tau\to -\infty$,
	or in both limits (if $|b| < h/2$). Again, it is finite as $\tau\to 0$
	($K_1=-1$).

\medskip\noi
	{\bf 4.} For $n=3$ ($D=6$), \Eq{v1} gives just $3k^2 = 2h^2$, therefore
	the solution behaviour is uniquely characterized in terms of the proper
	time $t$. Thus, for $K_1=0$ (spatially flat world) the expanding
	model has the aysmptotics (where arbitrary constants are chosen with no
	loss of generality)
\bear                                                       \label{v3}
	t \al\sim\al \e^{(-\sqrt{3/2}+1/4)h|\tau|}  \to 0,
                                                            \nn
	a(t) \al\sim\al \e^{(-\sqrt{1/6}+1/4} h |\tau| \sim t^{0.162}, \nn
	\e^{\beta_2} \al\sim\al \e^{-h|\tau|/2} \sim t^{0.514}
\ear
	--- initial singularity, and
\bear                                                         \label{v4}
	t \al\sim\al \e^{(\sqrt{3/2}+1/4)h|\tau|}  \to 0,
                                                                    \nn
	a(t) \al\sim\al \e^{(\sqrt{1/6}+1/4) h |\tau|} \sim t^{0.447}, \nn
	\e^{\beta_2} \al\sim\al \e^{-h|\tau|/2} \sim t^{-0.34}
\ear
	--- expansion in physical space and contraction in extra dimensions.

	For $K_1=+1$ (spherical world), the time symmetry of the evolution is
	broken only by the shift $\tau_2$ in $\e^z$ which does not affect the
	asymptotics. The evolution proceeds between two singularities like
	(\ref{v3}).

	For $K_1=-1$ (hyperbolic world) the model has a singularity like
	(\ref{v3}) at $\tau=\infty$; on the other hand,
\bear                                                         \label{v5}
	t    \al \sim \al  1/\sqrt{\tau} \to \infty; \nn
	a(t) \al \sim \al  1/\sqrt{t}\sim t \qquad {\rm as} \quad \tau\to 0,
\ear
	i.e. there is a linear expansion (contraction) asymptotic with the
	internal scale factor $\e^{\beta_2}$ tending to a finite limit.

\medskip\noi
	{\bf 5.} For $n > 3$, the advent of one more constant $b$ makes the
	evolution of $a(t)$ more diverse. One can observe, in particular, that,
	for any set of constants, there is always an asymptotic like (\ref{v5})
	for $K_1=-1$ and at least one singularity like (\ref{v3}) (with, in
	general, other numerical characteristics) for $K=+1$.

\medskip

	It should be stressed that the behaviour of $a(t)$ and even the
	finiteness or infiniteness of time intervals crucially depend on the
	choice of the conformal frame. Thus, the present vacuum solutions
	for $n=3$ coincide with those of \Ref{Fa3}, but the conclusions are
	different since there the authors considered the metric $g\mn$ as the
	physical one.

	In what follows we only indicate the integrable cases, postponing their
	detailed analysis to further publications.

\section{Solutions with matter: $K_1=0$}

	The unknown function $y$ is excluded from \Eq{c20} in two cases:
\begin{enumerate}
\item
	if $y$ is proportional to $x$ (i.e., to $\beta_2$),
\item
	if $A_2=0$.
\end{enumerate}

\noi	{\bf 1.} This possibility is realized if $a_1=a_3=2$ (stiff matter
     equation of state in $M_1$ and $M_3$. Then we have
\beq                                                               
	\ddot \beta_1 = \ddot \beta_3 = -\ddot \beta_2,      \label{c28}
\eeq
	so that it is sufficient to integrate (\ref{c20}) for $\beta_2$,
	then $\beta_1$ and $\beta_3$ will be easily found. \Eq{c20} reduces to
	the form
\bearr                                                             
	\ddot\beta_2 + Q^2\e^{2(n-1)\beta_2}
	             + (1-\half a_1) \e^{(n-1)(2-a_2)\beta_2}=0. \nnn
          			   							\label{c29}
\ear
	Thus it belongs to the type
\bearr                                                             
	\ddot x + B_1\e^{2x} + B_2 \e^{2cx}=0, \nnn
	\cm B_1,\ B_2,\ c = \const.                          \label{c30}
\ear
	Its first integral is
\beq                                                               
	\dot x{}^2 + B_1\e^{2x} + (B_2/c)\e^{2cx} = \const    \label{c31}
\eeq
	which is evidently integrable by quadratures.
	The solution is expressed in elementary functions in the cases
     $c = 1/2,\ 1,\ 2$. By (\ref{c29}) this corresponds to
\beq                                                               
	a_2 = 1,\ 0,\ -2,                                       \label{c32}
\eeq
	the first two values belonging to the physical range.

	So we have obtained a class of solutions which can be labelled
\beq
	(2n,0,F\mid 2,a_2,2),                                   \label{c33}
\eeq
	where the first three symbols designate the dimension, spatial
	curvature (in $M_1$) and a nonzero F field, while the remaining three
	indicate the coefficients $a_i$.

\medskip\noi
     {\bf 2.} The relation $A_2=0$ is explicitly written as
\beq
	2 + (n-1)a_2 = 3a_1 + (n-3)a_3.                        \label{c34}
\eeq
	So the equation for $\beta_2$ is integrated in the form (\ref{c24}).
	Then, if we combine \eqs (\ref{c19})--(\ref{c21}) to form
	$\ddot y$ in the left-hand side, it turns out that, in the right-hand
	side, the coefficient before $\e^{2x}$ vanishes and we obtain
\bearr
     \nhq	  2\ddot y =[3(2{-}a_1)(a_1{-}a_2)
	                + (n{-}3)(2{-}a_3)(a_3{-}a_2)]\e^{2y}\nnn
												 \label{c35}
\ear
	which is easily integrated like (\ref{c25}). So we know $\beta_2$ and
	$y$. Lastly, there is a linear combination of (\ref{c19})--(\ref{c21})
	whose right-hand side is zero, namely,
\beq
	(a_3-a_2) \ddot\beta_1 + (a_3-a_1) \ddot\beta_2 +
	                         (a_2-a_1) \ddot\beta_3 =0.     \label{c36}
\eeq
	In the general case \eqs (\ref{c35}) and (\ref{c36}) are independent,
	so after trivially solving (\ref{c36}) the integration is completed.
	In the exceptional case when they are not, one can find $\beta_1$ from
	(\ref{c19}), since both $x$ and $y$ are already known.

	Thus for $K_1=0$ we have obtained two families of solutions: (\ref{c33})
	(one-parameter) and
\beq
	(2n,0,F\mid a_1,a_2,a_3),                               \label{c37}
\eeq
	constrained by (\ref{c34}) (two-parameter). Their only common element
	is the trivial case of all $a_i=2$, when all $A_i=0$ and the solution
	is reduced to vacuum up to the relation between the integration
	constants that follows from the energy integral. This case of stiff
	matter in the whole space coincides with that of a minimally coupled
	scalar field $\varphi(\tau)$, to be discussed later.

\section{Solutions with matter: $K_1 = \pm 1$}

	The same combination of (\ref{c19})--(\ref{c21}) that has led to
	\Eq{c25} in the vacuum case, for $\rho_0\ne 0$ reads:
\beq                           \wider
	(\gamma-\beta_1)\ddot{} + 4K_1 \e^{2\gamma-2\beta_1}
					    = \kappa\rho_0 (2-a_1)\e^{2y}. \label{c38}
\eeq
	One can again select two cases of its integrability:
\begin{enumerate}
\item
	$a_1=2$ [then (\ref{c38}) coincides with (\ref{c25})] and
\item
	$y$ is proportional to $(\gamma-\beta_1)$ and (\ref{c38}) reduces to
	(\ref{c30}).
\end{enumerate}

\noi {\bf 1.} $a_1=2$, stiff matter in physical space. We are left with \eqs
     (\ref{c20}) and (\ref{c21}) with two unknowns $x$ and $y$. \Eq{c20} is
	easily integrated (reduces to (\ref{c24})) when $A_2=0$, that is
\beq
	(n-1) a_2 - (n-3) a_3 =4.                               \label{c39}
\eeq
	This condition is of interest only for $n>3$, since for $n=3$ it
	reduces to $a_2=2$, i.e., the trivial case $a_1=a_2=2$.

	For $n>3$, the function $y$ can be found from an equation similar to
	(\ref{c35}) but with another constant before $\e^{2y}$. This completes
	the integration.

	We thus obtain the 1-parameter family of models
\beq
	(2n,\pm 1, F\mid 2,a_2,a_3), \qquad  n>3	             \label{c40}
\eeq
	constrained by (\ref{c39}).

	For $a_1=2,\ n=3$, there is no $\beta_3$, and \Eq{c20} takes the form
\beq
	2\ddot\beta_2 = -2Q^2\e^{4\beta_2}
	               + \kappa\rho_0 (a_2-2)\e^{2\beta_2(2-a_2)} \label{c41}
\eeq
	belonging to the type (\ref{c30}). It is integrable in elementary
	functions for $a_2=1$ (no pressure) and $a_2=0$ (vacuum-like state).
	Thus we obtain the models
\beq
	(6,\pm 1, F\mid 2, a_2).                                 \label{c42}
\eeq

	Another case with an equation like (\ref{c41}) exists for $n>3$: if
	$a_1 = a_3 =2$, then
\beq
	2y = (n-1)(2-a_2)\beta_2.                                \label{c43}
\eeq
	Thus \Eq{c20} is solved; then, knowing the combination $\gamma-\beta_1$
	and $\beta_2$ (hence also $y$), it is straightforward to find $\beta_3$
	from (\ref{c21}). Thus we obtain a family of models similar to
	(\ref{c33})
\beq
	(2n, \pm 1, F\mid 2, a_2, 2), \qquad n>3.                \label{c44}
\eeq

\medskip\noi
	{\bf 2.} $a_1\ne 2$;  to solve (\ref{c38}), let us require that
	the two 3-vectors of coefficients by $\beta_i$ in the expressions for
	$y$ and $\gamma-\beta_1$ be parallel:
\beq
	\pmatrix{ 3(2-a_1)\cr
			(n-1)(2-a_2)\cr
			(n-3)(2-a_3)\cr }         \sim
	\pmatrix{ 2 \cr n-1 \cr n-3 \cr } ,                    \label{c45}
\eeq
	whence
\beq
	a_2 = a_3 = \fract 32 a_1 -1.                           \label{c46}
\eeq
	In this case also automatically $A_2=A_3=0$, so that $\beta_2$ and
	$\beta_3$ are found quite easily, thus completing the integration.
	We get the models
\beq
	(2n,\pm 1, F \mid a_1, \fract{3}{2}a_1-1, \fract{3}{2}a_1-1).
												  \label{c47}
\eeq

	Thus we have obtained 4 one-parameter families of models with nonzero
	spatial curvature. In each case, to obtain a final form of the
	solution, it is necessary to substitute the results of integration to
	the first integral (\ref{c23}) to find a relation between the
	integration constants.

\section{Inclusion of a massless scalar field}

	So far we have been working with the action (\ref{a}). However, all the
	above solutions can easily include a massless, minimally coupled
	scalar field $\varphi=\phim$ with the action
\beq
	S[\phim] = \int \sqrt{g} (\nabla \varphi)^2.             \label{5.2}
\eeq
	Then, in the space-time considered, the scalar field equation for
	$\varphi=\varphi(\tau)$ is immediately integrated to give
\beq
	\varphi = \varphi_0 + \varphi_1\tau; \cm \varphi_0,\varphi_1=\const.
												  \label{5.3}
\eeq
	The EMT of \phim\ coincides up to a constant factor with that of stiff
	matter (when all $a_i=2$).

	It is easily verified that \phim\ does not contribute to the
	right-hand side of \eqs(\ref{c10}) and therefore in no way affects the
	solution process described above. It only affects the system through
	its contribution to the first integral (\ref{c23}) where one should add
	the term $\kappa\varphi_1^2$ to the right-hand side. The same applies
	if one adds one more component of matter in the form of a perfect fluid
	with stiff equation of state ($a_1=2$): then the addition is
	just $\kappa\rho_{02} = \const$. This addition certainly changes some
	properties of the solutions.

	Due to the conformal invariance of the $F$ field,
	$D$-dimensional conformal transformations \cite{wag,br-birk}
	lead to models of generalized scalar-tensor theories of gravity. Vacuum
	solutions are then generalized in a straightforward way, whereas
	perfect fluids will in general acquire a nontrivial interaction with
	scalar fields.

\Acknow
	{This work was supported in part by the Russian State Committee for
	Science and Technology and Russian Basic Research Foundation.

	I am grateful to Julio Fabris for numerous helpful discussions
	and to CNPq, Brazil, for financial support during my stay at
	UFES, Brazil, where part of the work was done.}

\end{document}